\documentclass{aastex}          
\usepackage{graphics}
\usepackage{graphicx}
\usepackage{dcolumn}

\newcommand{\be}{\begin{equation}}
\newcommand{\ee}{\end{equation}}
\newcommand{\ba}{\begin{eqnarray}}
\newcommand{\ea}{\end{eqnarray}}

\begin{document}
\title{Bragg diffraction and the Iron crust of Cold Neutron Stars}

\shorttitle{Bragg diffraction off Neutron Stars}
\shortauthors{Llanes-Estrada and Moreno Navarro}

\author{Felipe J. Llanes-Estrada and Gaspar Moreno Navarro} 
\affil{Departamento de F\'{\i}sica Te\'orica I, Universidad Complutense, 28040 Madrid, Spain}
\email{fllanes@fis.ucm.es} 

\begin{abstract}
If cooled-down neutron stars have a thin atomic crystalline--iron crust,
they must diffract X-rays of appropriate wavelength. If the diffracted beam is to be visible from Earth (an extremely rare but possible situation), the illuminating source must be very intense and near the reflecting star. An example is a binary system composed of two neutron stars in close orbit, one of them inert, the other an X-ray pulsar\footnote{Perhaps an ``anomalous'' X-ray pulsar or magnetar, not powered by gas absorption from the companion or surrounding space, would be the cleanest example}. 
The observable to be searched for is a secondary peak added (quasi-) periodically to the main X-ray pulse. The distinguishing feature of this secondary peak is that it appears at wavelengths related by simple integer numbers, $\lambda$, $\lambda/2$, $\lambda/3$... $\lambda/n$  because of  Bragg's diffraction law.
\end{abstract}
\keywords{Bragg difraction, neutron stars, crystalline iron, Bragg peaks}

\section{Introduction}

Current theories of Neutron Stars~\citep{Bombaci:2007zz} imply that as
pressure builds up towards the interior of the star there are successive
phase transitions from an iron crust~\citep{camel,deYoung}, to a nuclear
medium with high neutron density, to a neutron Fermi liquid, to more exotic
forms of matter. It has proven difficult to make empirical
progress on the star's composition. For example, very information-rich
equations of state~\citep{Huber:1994zz,Klahn:2006ir} have to be tested with
few numbers (mass, size, pulsar period and time dependence, etc.)

That the outer-most layer of the star contains iron is known from the
characteristic absorption lines of Fe~\citep{Xabsorption}. 
In recently formed stars or in stars that are heated by accretion, a hot iron ``ocean'' is likely to cover the surface \citep{Yakovlev}. However, if the star is not accreting material, it cools rapidly; the star stops emitting X-rays after about a million years,  \citep{camel}. At about $2\times 10^7$ years (a small time in galactic scale) the star is believed to have reached $1000\ K$ simply by radiation \citep{Yakopethick,cooling}, a temperature which is well below the solidification of iron.

The hypothesis that this iron layer is in crystalline form has only been indirectly tested by observations of the initial cooling in
quasi-persistent soft X-ray transients. These seem to be consistent with the neutron star crust having the structure of a perfect crystal, while models based on an amorphous crust cannot fit the data~\citep{Shternin}.

We here propose a direct test that might be performed as the catalog of X-ray sources expands, requiring the existence of a binary system composed of one such cooled-down neutron star with certain minimal assumptions that will be spelled out shortly, and an X-ray emitting companion.

The canonical way of ascertaining the crystalline structure in a laboratory
material is by exposing it to X-rays and studying the resulting diffraction
pattern (it is not possible to directly observe neutron diffraction from Earth due to the neutron's short lifetime of about 15min). This textbook method
works well even for fragmented crystals, for example a dust sample made of
microcrystals can be made to produce a diffraction pattern that appears as characteristic diffraction rings~\citep{cullity}.

The concept is the same in an astronomical context: the \textquotedblleft
sample\textquotedblright\ is the neutron star whose crystalline crust
is being examined, and one needs a beam powerful enough to detect it, after
reflection on the star, in Earth or in a orbiting X-ray telescope. This
stringent condition requires a very intense beam to be focused on the star
and then reflected onto our direction.

A promising and simple system that offers a potential opportunity to explore
this diffraction is a binary system composed of two neutron stars. We will
assume one of them, nicknamed \textquotedblleft \emph{Pharus}%
\textquotedblright , to be an active X-ray pulsar directly visible from
Earth. Many pulsars are known to emit in X-rays \citep{Campana:2009bz}. Its
period can be taken of order 1-second, as typical for the larger part of
the normal pulsar population. Its companion in the binary system, that we
will name \textquotedblleft \emph{Reflector}\textquotedblright\ is assumed to
be another compact object, indeed a neutron star, but this one dark and
inert, held by its neutron degeneracy pressure but emitting no significant
radiation, nor possessing a strong magnetic field that may alter the reflection of the beam from Pharus. 
If the orbital conditions are right, then one may observe the direct pulses
from Pharus, but also at least one diffracted beam from Reflector, depending
on the frequency of X-ray observation.

A further complication that may arise is
if the neutron star possesses an atmosphere of ionized gas
above the thin iron crust that we would like to see detected. This
possibility has been analyzed with quite some care in~\citep{Suleimanov:2009wx}. In fact, part of the X-rays in a magnetized star
are scattered or absorbed by this atmosphere. It is not clear whether this
is a general phenomenon and applies also to a non-magnetic star, but it nonetheless needs to be kept in mind. A double (binary) pulsar would be more
difficult to analyze since in addition to the main and reflected pulses
there would be a second main pulse set and maybe even another set of secondaries.
Moreover the strong magnetic field of the mirror star could distort the
reflecting crystal.

Of course many other forms of intense X-ray emission are possible, such as
bursts from accretion processes in X-ray binaries or from black-hole or
supernovae ejecta~\citep{Borkowski:2007nk}. These should also provide the
equivalent of Bragg peaks (but without the characteristic main pulse of
Pharus).
An advantage of a binary neutron--star system is that the main pulse
provides a clock--time $T^{\mathrm{spin}}$ that is useful in the data
analysis.
However the primary emitter could also obscure the signal if its thermal emission would be significantly larger than the secondary reflections. 
This is not the case for younger pulsars such as Crab or 0540-69~\citep{Kunzl}, that show no significant thermal emission, but might erase our proposed signal for older pulsars in the soft-X ray region that includes the first two Bragg peaks off iron. Other peaks, in the harder X-ray region, are less likely to be masked by possible bremsstrahlung backgrounds.

\section{Orbital conditions for observability of Reflector}
We study a simple orbital case, in which the two stars have equal
mass and are therefore equidistant to the center of mass. This is an
extreme situation and, on a case by case basis, one could study more
general configurations should data become available. Known pulsars forming
part of double neutron stars have masses that cluster around 1.35 solar
masses with a spread of order 0.04 solar masses \citep{Thorsett:1998uc}.
Therefore this case is close to what one might expect
to find when looking for a double neutron--star system.

\begin{figure}[htbp]
\includegraphics[width=5cm]{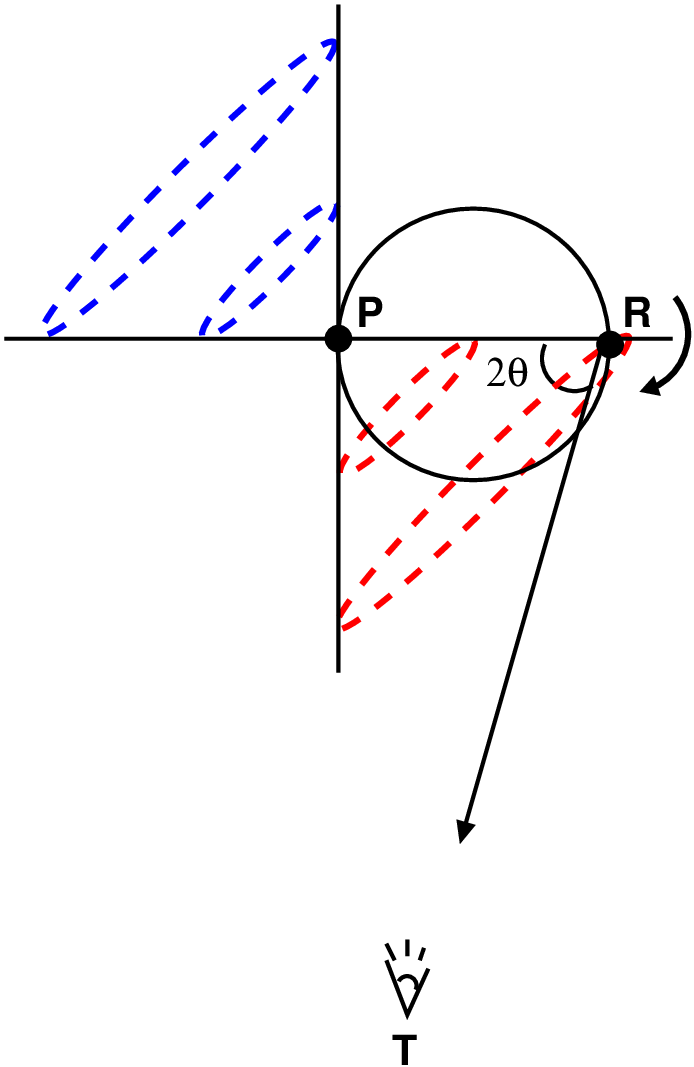}
\caption{T denotes the terrestrial observer; P the pulsating star Pharus; R
the reflecting star Reflector; if R in its orbital motion crosses the cone
illuminated by P's X-ray beam at a time where the crossing is illuminated,
then Bragg diffraction might occur, visible at a wavelength meeting the Bragg condition for the reflecting angle $\protect\theta$. }
\label{fig:orbitS}
\end{figure}

Pharus presents, as typical for pulsars, a strong magnetic field precessing
around its spin axis. X-rays are emitted along the direction of the magnetic
field lines streaming out the magnetic pole; thus, the radiation beam,
which is observed as a time series of flashes as it rotates, swipes a conic
surface. The visual to Earth, PT in figure \ref{fig:orbitS}, has to be contained in this surface, since, after all, Pharus is assumed to be observed as a conventional X-ray pulsar.

Further, we will assume that the plane of the orbit of Reflector (taken
circular) contains the visual to Earth, so that the orbit is seen on
edge. These are not necessary conditions, and a computer programme can be
written to incorporate more general cases if putative data becomes
available. The situation is depicted in figure \ref{fig:orbitS}.

The beam cone intersects the orbit of Reflector in at least one point,  and its
angular position along the orbit will depend on the opening of Pharus's cone
(angular width), controlled by its magnetic anomaly or inclination angle to
the rotation axis. 

If Reflector happens to be at the orbit-beam cone crossing when Pharus
illuminates it, the reflected radiation might reach Earth at an instant when
the pulsar is, in theory, still.

In our simple computer simulation, the intensity of the radiation of Pharus $I_P$ for a given X-ray wavelength will be taken as the unity at maximum, with a Gaussian profile
(this plays no role as long as it is a simple function). The intensity of
the beam diffracted by reflector $I_R$ will naturally be smaller. This will be partly due to
absorption in Reflector, but primarily because the illumination of its
surface will decrease with the square of the distance to Pharus, 
\begin{equation}
I_{R}=I_{P}\frac{r^{2}}{R^{2}}
\end{equation}%
where $r$ is the effective emitting radius of Pharus (of order 100 km
for a typical neutron-star pulsar of radius 10 km, as estimated from the magnetosphere's end at the light cylinder~\citep{Saito}) and $R$ the radius of the
orbit of Reflector.
A further decrease of the reflected signal is brought about by the fraction of the surface that reflects Pharus's beam. This, in analogy with the lunar phases, is
$$
F=\frac{1}{2}\left( 1+\cos 2\theta\right)
$$   
and is incorporated in the example below.

To keep the signal of Reflector within one or few percent of the intensity of
Pharus, Reflector has to orbit in close range to Pharus. However, to have a
clearer line-shape for the signal it is convenient to have the orbital
period of Reflector, $T^{\mathrm{orbit}}$ to be of the same order as, (but
larger than) the spinning period of Pharus, $T^{\mathrm{spin}}\simeq 1s$,
which demands, because of Kepler's third law, and in terms of the two star's masses and Cavendish's constant $G$,
\[
T^{\mathrm{orbit}}=\frac{2\pi R^{3/2}}{\sqrt{(M_P+M_R)G}}=\frac{2\pi
R^{3/2}}{\sqrt{2M_RG}}\ \ \mathrm{if}\ \ M_P=M_R
\]
larger orbital radii. Given the tension between the two requirements, we
compromise to an orbit of order 10000 km (or a period $T^{\mathrm orbit}$ of order $15\ s$), which is certainly small enough for
relativistic corrections to the orbit to play a role, but we will ignore
these. (Note that the double neutron star binary  with the
shortest orbital period is, as of 2004, 2.45 hours in J0737--3030 \citep{Lyne2004}, so that this case study is  low compared with current
observations of binary systems, but not impossible).

To check that relativistic corrections are not a major source of uncertainty, we take  the relevant equations of General Relativity~\citep{Peters1964} to estimate in turn the power radiated by
gravitational waves, the rate of decrease of the semi major axis, the rate
of decrease of the orbital period, and the total time for collapse of the
binary, yielding, for typical values of the star's mass and the period/orbital radius just quoted, 

\begin{eqnarray}
P =\frac{32}{5}\frac{G^{4}M_P^{2}M_R^{2}(M_P+M_R)}{c^{5}R^5} \sim 10^{35} \rm{Watt} \\
-\frac{dR}{dt} =\frac{64}{5}\frac{2 G^{3}M_P^{3}}{c^{5}R^{3}}
 \sim 0.01-0.1\rm{ m/s} \\
-\frac{dT^{orbit}}{dt} =\\ \nonumber \frac{96}{5}\frac{G^{3}M_PM_R(M_P+M_R)}{%
c^{5}(T^{\rm orbit})^{\frac{5}{3}}}(\frac{4\pi ^{2}}{G(M_P+M_R)})^{\frac{4}{3}%
}\sim \\ \nonumber
1-10\mu \rm{s/year} \\
T^{\rm{collapse}} \sim 5-10 \rm{years}
\end{eqnarray}

Thus, relativistic effects may be neglected if the
observation time of pulsar profiles is very short  compared with the
time to collapse just estimated.

The extraction of the reflection angle for a given binary system 
might employ the variation of Pharus' velocity along the line of sight in its orbital motion, causing a first--order kinematic Doppler shift, affecting the spectral shape of the main peak. Analyzing the red/blue shift of the peak over several periods one can determine from the data to what angular position along the orbit does a given main peak of Pharus correspond. Then one just needs to note that the position of Reflector a short time $\Delta t$ later takes a correction
\be
\theta = \theta_0^{\rm Doppler} + \frac{2\pi \Delta t}{T^{\rm   period}} \ .
\ee


\section{Bragg diffraction at Reflector}

\begin{figure}[htbp]
\includegraphics[width=5cm]{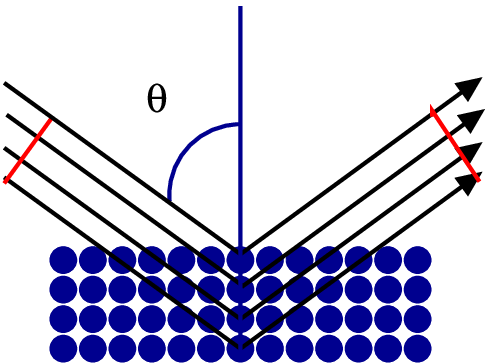}
\caption{If the radiation's wavelength  is comparable to the crystal
spacing, the crystal acts as a diffraction grating, reflecting different
wavelengths by different angles. The phenomenon is understood as
constructive interference,with the rays (black arrows) having all equal
phase at the wave front (transverse red lines). The additional path
travelled by rays penetrating deeper into the crystal needs to be an integer
number of wavelengths for the phase of these rays to match those reflected
at the surface.  Bragg's law follows. }
\label{fig:Bragg}
\end{figure}

The principle of Bragg diffraction is illustrated in figure \ref{fig:Bragg}. 
A crystalline material acts as a difraction grating. A beam of a given
wavelength is reflected, by constructive interference, only when Bragg's law
is met 
\begin{equation}  \label{Braggslaw}
\sin (\frac{\pi}{2}-\theta) = \cos\theta = \frac{n\lambda}{2d} \ ,
\end{equation}
with $d$ the spacing between crystal planes, $\lambda$ the radiation
wavelength, $\theta$ the reflection angle taken from the vertical, and $n$
an integer number. Because the sine is smaller than one and $n$ greater, this
condition can only be satisfied for wavelengths smaller than twice the
crystal spacing.

It is known that atomic iron crystals present a body-centered cubic
structure with spacing $d=0.3nm$ called ferrite at temperatures above $1665\ K$ or below $1184\ K$. Austenite, a crystallization in a face-centered cubic lattice is stable between these two temperatures \citep{Kawasaki} .
These cubic crystallizations  are already quite optimal and
therefore we can assume that the outermost layer of a neutron star, if
crystalline iron, will employ them depending on the temperature. 
Therefore, if the
spacing $d$ is not altered in neutron stars, the characteristic wavelengths that need to be scanned are $\lambda <0.6nm$.

That the lattice spacing remains, at the neutron star's surface, similar to that under earthly conditions requires closer examination. First, we show
from a macroscopic point of view that gravitational forces at the very surface of Reflector only moderately deform the crystal. Indeed,
the inwards gravitational acceleration  at the surface is of
order $g= 10^{12}m/s^{2}$. This causes a force per unit area $F/A_0=g\rho h$ to depth $h$ inside the crust (the density of iron being $\rho=7.87\ g/cm^3$).
Given the Young modulus of iron, $Y=196.5 GPa$, the distortion of the lattice expected is $\Delta d/d = F/(Y\ A_0)$, which grows linearly with depth.
The penetration depth of X-rays depends on their frequency and is typically $2-15\mu m$ in crystalline iron \citep{Randle}(this amounts to a number of about $10^{4}-10^{5}$ planes necessary to totally diffract the beam). $\Delta d/d$ remains significantly smaller than 1 for all layers within this penetration depth.

Another argument is to consider that characteristic microscopic
atomic forces are of order $2.5 \times 10^{-8}$ N and these, based in
Coulomb interactions between electrons and iron nuclei, are also much
stronger than gravitational forces exerted on the atomic mass, of order $%
3\times 10^{-12}$ N. 
Thus, for the purposes of X-ray diffraction, crystalline iron at the star's surface is similar to that on Earth, with somewhat broadened lines.
This is in spite of the increasingly large deformation that happens with depth. Deeper in the crust densities are enormous, easily two orders of magnitude larger than in Earth, so that the atomic crystal is deformed and diffraction occurs at lower wavelengths. However such depths are hard to reach for X-rays that diffract near the surface. 

However, the radiation wavelength that we observe is red-shifted respect to
the radiation reflected at the star's surface, due to the gravitational
potential in General Relativity. This effect brings about a correction
necessary when extracting the crystal spacing from Bragg's law, since the $
\lambda $ appearing in that formula is taken at the star's surface. One
finds a gravitational red shift $z$ 
\be
\lambda ^{\mathrm{Earth}}=\lambda (1+z)= \frac{\lambda}{\sqrt{1-2M_RG/Rc^{2}}}
\ee
in terms of the star radius and mass. (This correction factor
is about 1.3 for a typical neutron star).

Equation (\ref{Braggslaw}) would then be corrected due to the gravitational red--shift to 
\begin{equation}  \label{Braggslaw2}
\cos\theta = \frac{n\lambda^{\mathrm{Earth}}}{2d(1+z)} \ .
\end{equation}
This formula can also be corrected for kinematic Doppler shift due to the orbital motion of Reflector as neeed be, simply substituting the factor $(1+z)$.

If a diffraction pattern was suspected, one could conceivably identify the form of crystallization (whether ferrite or austenite) by the lines present. This is possible because the fcc and bcc crystal have different symmetry planes.
In terms of the Miller indices $h$, $k$, $l$ and the unit-cell size $a$, the spacing between those planes is given by
\be
d^2= \frac{a^2}{h^2+k^2+l^2}\ , 
\ee
and combining with Eq.(\ref{Braggslaw2})
\be
\cos \theta = \frac{n\lambda^{\mathrm{Earth}}\sqrt{h^2+k^2+l^2}}{2a(1+z)} \ .
\ee
For a bcc lattice only those combinations where $h+k+l$ is an even integer occur in the diffraction pattern. An fcc lattice diffracts instead for $h,k,l$ all odd or all even.
\begin{table}
\caption{\label{tablita:fccbcc}Inverse of the interplane spacing relative to the unit-cell size $a/d$ and multiplicity factor $P$ (how many combinations of Miller indices lead to the same $d$) for the fcc (left) and bcc(right) lattices.}
\begin{tabular}{|ccc|ccc|}
\hline
$hkl$ & $\sqrt{h^2+k^2+l^2}$ & P & $hkl$ & $\sqrt{h^2+k^2+l^2}$ & P \\
\hline
111 & 1.73 & 8   & 011 & 1.41 & 12  \\
\hline
002 & 2    & 6   & 002 & 2    & 6   \\
\hline
    &      &     & 112 & 2.45 & 24  \\
\hline
022 & 2.83 & 12  & 022 & 2.83 & 12  \\
\hline
113 & 3.32 & 24  & 013 & 3.16 & 24  \\
\hline
222 & 3.46 & 8   & 222 & 3.46 & 8   \\
\hline
    &      &     & 123 & 3.74 & 48  \\
\hline
004 & 4    & 6   & 004 & 4    & 6   \\
\hline
331 & 4.36 & 24  & 411 & 4.24 & 24  \\
\hline
024 & 4.47 & 24  & 024 & 4.47 & 24  \\
\hline
    &      &     & 233 & 4.69 & 24  \\
\hline
224 & 4.9  & 24  & 224 & 4.9  & 24  \\
\hline
333 & 5.2  & 32  & 105 & 5.1  & 72  \\
511 &      &     & 314 &      &     \\
\hline
    &      &     & 215 & 5.48 & 48  \\
\hline
044 & 5.66 & 12  & 044 & 5.66 & 12  \\
\hline
135 & 5.92 & 48  &     &      &     \\
\hline
    &      &     & 035 & 5.83 & 48  \\
    &      &     & 334 &      &     \\
\hline 
424 & 6    & 30  & 424 & 6    & 30  \\
006 &      &     & 006 &      &     \\
\hline
\end{tabular}
\end{table}
Table \ref{tablita:fccbcc} shows the first few diffraction lines expected. Given are the Miller indices, the distance between diffracting planes with given Miller indices in lattice units, and the multiplicity of planes yielding the same distance (and thus diffracting at the same wavelength), the intensity of the diffracted beam being proportional to this multiplicity.
Additional factors influencing the intensity of a given line are the Lorentz-polarization factor and the atomic scattering factor for Fe, given in 
\citep{cullity}.


\section{Line shape of the pulsar}

\vspace{0.5cm}

\begin{figure}[htbp]
\includegraphics[width=8cm]{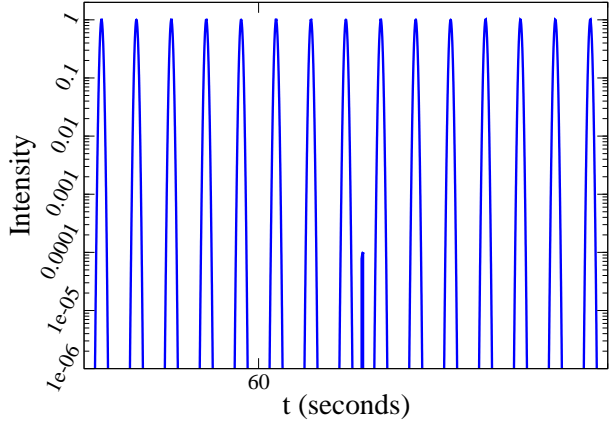}
\caption{The spectrum of the binary system with $M_R=M_P$. for several revolution periods of the system (about 15 seconds each). 
The intensity
of Reflector (smaller peak in the center of the figure) relative to Pharus (taken as 1 at its maximum) depends on the
albedo of Reflector as well as on its distance to Pharus and orbital phase.
}
\label{fig:lineshape2}
\end{figure}

An example of an observable line-shape for the binary X-ray pulsar is plotted in
figure \ref{fig:lineshape2}. 

One sees that, in addition to the pulse of Pharus, a smaller
signal due to Reflector appears. The intensity of this pulse depends on the
distance between Pharus and Reflector, but Kepler's third law links this to
the orbital period, and thus the relative intensity of the secondary pulse
is not independent of its repetition rate.

\begin{figure}[htbp]
\includegraphics[width=8cm]{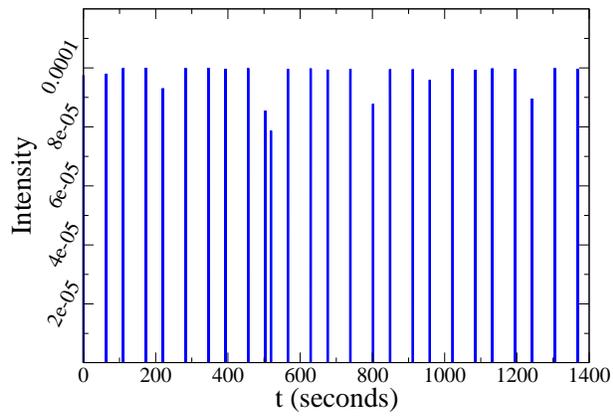}
\caption{Same as in figure \protect\ref{fig:lineshape2} but subtracting the main pulse of pharus to see several repetitions of the Reflector pulse.
The repetition rate depends on the orbital period of
Reflector and the spinning (pulsar) period of Pharus, since Reflector has to
cross Pharus's beam cone at a moment when this is illuminated. 
 Note that not all pulses have the same height due to the different illumination of Reflector in different crossings of the radiation cone of Pharus, and the orbital phase (fraction of illuminated surface visible from Earth). }
\label{fig:lineshape3}
\end{figure}

This periodic reapparition of Reflector every $\Delta t$ depends on the orbital coincidence
that it crosses the cone subtended by Pharus's beam when it is irradiated.
This happens when, for two integers $l_{1}$ and $l_{2}$, 
\begin{equation}
\Delta t=l_{1}T^{\mathrm{orbit}}=l_{2}T^{\mathrm{spin}}\ .
\end{equation}%
This rational number relation does not need to be exact, in view of the
time-width of Pharus's beam. 
We take some ad--hoc values for illustrational purposes that are $%
T_{orbit}\simeq 15\ sec$ and $T_{spin}=1\ sec$. The resulting line shape,
where the secondary is clearly observed, is represented in figure \ref%
{fig:lineshape2}. Note the secondary misses several periodic apparitions
since the spin and orbital period are not perfectly matched (which we expect
to be borne in reality). In figure \ref{fig:lineshape3} we represent the
same data for shorter periods so one can discernt the Gaussian shape of the
main pulse.

What distinguishes Bragg diffraction from other astrophysical phenomena that
produce complicated secondaries (indeed pulsars have rather varying
line-shapes \citep{Smith2003} is that the peak appears at very
specific wavelengths. Since $n$ in eq. (\ref{Braggslaw}) is an integer, the
secondary appears at the wavelength sequence 
\[
\lambda ,\ \frac{\lambda }{2},\ \frac{\lambda }{3},\ \frac{\lambda }{4}\dots 
\]%
and at other specific frequencies as dictated by the various planes in table
\ref{tablita:fccbcc}.

In addition,  for the degenerate case (orbit seen edge-on) that we have chosen,
there is one additional crossing of Pharus's beam cone with Reflector's
orbit. The reflection angles appearing in Bragg's law for these crossings
satisfy, given that the binary system subtends a negligible angle as seen
from Earth, $2\theta_1+2\theta_2\simeq \pi$. Then one can easily see that a
second series of wavelengths 
\[
\lambda^{\prime },\ \frac{\lambda^{\prime }}{2}, \ \frac{\lambda^{\prime }}{3%
},\ \frac{\lambda^{\prime }}{4} \dots 
\]
with $\lambda^{^{\prime }2}=(2d)^2-\lambda^2$ will also show a secondary
peak (same for other sequences caused by additional reflecting planes).

Another very interesting degenerate case occurs when the magnetic axis of
Pharus is perpendicular to its precession axis: then the X-ray beam swipes
the entire orbit of Reflector, and yields the possibility of detecting, at
short X-ray wavelengths, several diffraction maxima.
\vspace{0.6cm}

\section{Prospects for detection}

Detection of such binary system in our galactic neighboorhood would be a not too-difficult task. For example, consider the well-studied Crab pulsar at a distance of some $2\ kpc$. A couple of days ($10^5$ seconds) of data taken in 2001~\citep{Tennant2001}
yield 0.19 counts per second, mostly in the $1-2\ keV$ region, corresponding to a fraction of a nanometer and ideally suited for diffraction off iron.

Should the Crab pulsar have had a close-orbiting companion when these data were taken (we obviously do not believe this to be the case for many other reasons, among them that the violent merger would have occurred in the intervening decade), it could probably have been detected with the existing satellite missions Chandra and XMM-Newton. To see it, just notice that between the maximum sensitivity at the peak of the spectrum~\citep{Tennant2001} and the high energy tail where one runs out of counts, there are three orders of magnitude. This means that the instrument should have been able to find an intensity ratio of Reflector to Pharus of 
$I_R/I_P\simeq 10^{-3}$. With a two to three week observation time, the requisite $10^{-4}$ sensitivity would have been reached by the increased statistics. 

The same sensitivity can be found in other published work, for example~\citep{Weisskopf2004}. Observations of the Crab, Vela, Geminga and other pulsars concur in yielding three decades of sensitivity in spectral structure between maximum and minimum sensitivity.

However the likelihood that a merger happens within the galaxy is relatively small during a human lifetime. The rate of coalescence of two-neutron star binary systems
has been estimated in several works \citep{Kim2003,Regimbau2005}, given its interest for the gravitational wave detection programme at LIGO and VIRGO, as well as possible upgrades of these machines.

Estimates for double neutron star mergers inside our galaxy are (in order of magnitude) $10^{-5}\ yr^{-1}$. To obtain a good rate that allows a potential detection in, for example, a decade of operation, one needs to be able to detect the X-ray source to a distance of up to 30-40 $Mpc$ (more than a hundred times the galactic size) to include many other star formations of the local group. This is the goal for the gravitational wave detectors.

The detection of X-ray diffraction in a binary system would point out to the tightness of the orbit and the temporal approaching of its final collision. Therefore it would be interesting to explore, in future work, the possibility of employing X-ray diffraction as an early-warning system for those detectors.


\section{Conclusions and outlook}


In conclusion, we have attempted to develop an astronomical observable able
to test the hypothesis of a crystalline iron crust in cold neutron stars. The
search we suggest is for a secondary pulse in an X-ray pulsar indicating a
binary system. The tell-tale of crystalline structure is the appearance of
this secondary for a specific sequence of wavelengths $\lambda,\ \frac{%
\lambda}{2}, \ \frac{\lambda}{3},\ \frac{\lambda}{4} \dots $
Several such frequency sequences may appear due to the various crystal planes
and to the possibility of observing Reflector at more than one angle along the orbit. 

The phenomenon is quite the opposite of an eclipsing binary. Instead, here
the companion star is dark most of the time and only brightens when Bragg's
diffraction condition is satisfied.

To extract the crystal spacing $d$ from data one could
follow the (simplified) steps

\begin{itemize}
\item Detect a potential binary system with the proposed pattern. Since the absolute calibration of the energy is not known a priori due to various red shifts, one way to proceed would be to study the intensity cross correlations 
$\langle I(E) I(2E)\rangle$  as a function of energy.

\item Measure the period of Pharus's main pulse, $T^{\mathrm{spin}}$.

\item Perform a simple rational number analysis to obtain an approximate $T^{%
\mathrm{orbit}}$ for Reflector.

\item Obtain the angle $\theta$ by a Doppler measurement or possibly a direct orbital reconstruction.

\item Given $\lambda_0$, the largest wavelength where Reflector is seen, after
any corrections needed, use Bragg's law to obtain $d$.
\end{itemize}

X-ray diffraction might also occur at the surface of a cold white-dwarf acting as Reflector. A difference between such stars and neutron stars is that nuclear fusion chains are generally thought not to be complete in the progenitor star, one does not expect to find iron but a mixture of lighter elements such as oxygen, neon, silicon, etc. It would surely be interesting and complicated to examine any data that may become available. Immediately one notices that the observation of the iron crust on a cold neutron star is an additional test of the efficiency of fusion in its very massive progenitor (believed to be completely efficient).

Finally, as with many other theoretical proposals in high-energy as well as astrophysics,  discovery with present instrumentation will involve a good amount of serendipity, although ``accidental'' discoveries occur, while serious exclusion bounds that would put in question the existence of a crystalline crust are much more difficult and should await a future generation X-ray satellite.

\acknowledgments
The authors thank Krysty Dyer and Katja Waidelich for a careful reading of the first manuscript.
This work was supported in part by grants FPA2007-29115-E, MCYT FPA
2008-00592/FPA, FIS2008-01323 (Spain).

\end{document}